\documentstyle[12pt]{article}

\newcommand{\be}{\begin{equation}}
\newcommand{\ee}{\end{equation}}
\newcommand{\bea}{\begin{array}}
\newcommand{\ea}{\end{array}}
\newcommand{\beqa}{\begin{eqnarray}}
\newcommand{\eeqa}{\end{eqnarray}}
\newcommand{\bean}{\begin{eqnarray*}}
\newcommand{\eean}{\end{eqnarray*}}

\def\up#1{\leavevmode \raise.16ex\hbox{#1}}

\def\sqr#1#2{{\vcenter{\vbox{\hrule height.#2pt
        \hbox{\vrule width.#2pt height#1pt \kern#1pt
          \vrule width.#2pt}
        \hrule height.#2pt}}}}

\newcommand{\journal}[4]{{\sl #1 }{\bf #2} \up(19#3\up) #4}

\newcommand{\gapproxeq}{\lower .7ex\hbox{$\;\stackrel{\textstyle
>}{\sim}\;$}}
\newcommand{\lapproxeq}{\lower .7ex\hbox{$\;\stackrel{\textstyle
<}{\sim}\;$}}

\newcounter{appendice}

\def\thebibliography#1{{\bf REFERENCES\markboth
 {REFERENCES}{REFERENCES}}\list
 {[\arabic{enumi}]}{\settowidth\labelwidth{[#1]}\leftmargin\labelwidth
 \advance\leftmargin\labelsep
 \usecounter{enumi}}
 \def\newblock{\hskip .11em plus .33em minus -.07em}
 \sloppy
 \sfcode`\.=1000\relax}

\begin{document}
\title{\hfill $\mbox{\small{
$\stackrel{\rm\textstyle } {\rm\textstyle
hep-th/0002229\quad\quad} $}}$ \\[1truecm]
   Dual Instantons   }
\author{A. Pinzul and A. Stern}
  \maketitle
 \thispagestyle{empty}

\begin{center}
 { Department of Physics, University of Alabama,\\
Tuscaloosa, AL 35487, USA.}
\end{center}

\begin{abstract}

We show how to map the Belavin-Polyakov instantons
 of the $O(3)$-nonlinear  $\sigma-$model
  to a dual theory where they then appear as  nontopological
solitons.   They are stationary points of the Euclidean action in the dual
theory, and moreover, the dual action and
the $O(3)$-nonlinear  $\sigma-$model action agree on shell.

\end{abstract}

\bigskip



\newpage

Although many techniques have now been devised
for finding dual  descriptions of field theories, important questions
and limitations remain.  (For reviews see\cite{rtd}.)
 One  limitation is that most
  of the techniques are applicable only in two space-time dimensions.
 Within the realm of nonabelian T-duality, there
  are issues concerning  the global aspects of the theory.
T-dual theories are  equivalent at the level
of the classical dynamics, and also
to several orders in perturbation theory.
Moreover,  from the current algebras,
the  dual descriptions are known to be   canonically
 equivalent.
But canonical equivalence is insufficient for proving the
equivalence of Feynman path integrals.\cite{GJ}
More troubling is the result that the canonical equivalence
between dual theories  is, in general,  only valid  locally,
as the configuration spaces of the theories can have
different   topological properties.    We can  have that certain
solutions are `topological' in one theory, but not in its dual,
as we shall demonstrate  here.  Could this
 lead to a breakdown in the quantum equivalence of the two theories?

The example we shall look at is that of the
$O(3)$ nonlinear $\sigma-$model,
having a   target  space of $S^2$.  From the condition of finite
action in two-dimensional Euclidean space-time, one gets that the
configuration  space is a union of disjoint pieces, and well known
topological  solitons appear, namely
 the Belavin-Polyakov  instantons.\cite{BP}
  The Bogomol'nyi bound insures that these solutions
are the minima in every topological sector of the theory.\cite{Bog}
Recently, a Lagrangian and Hamiltonian field theory  was  found which
 is {\it locally}  canonically equivalent to
the $O(3)$ nonlinear $\sigma-$model, and
is  generalizable to nonlinear $G/H$ models
for any Lie groups $G$ and $H\subset G$.\cite{prwo}
\footnote{It is also generalizable to dynamics consistent with
Poisson-Lie T-duality\cite{KS}, \cite{Sf},\cite{prwo}.}   However,
the target space for this dual theory
is topologically trivial, and finite action restrictions
do not lead to any disconnected regions of the configuration
space.  (Of course, at the classical level
 the dual action can only be determined up to divergence terms
 since one only demands equivalence of equations of motion.
But there are no divergence terms that can be added to the
 dual Euclidean action for the  purpose of obtaining a nontrivial
 topology, and moreover the dual Euclidean action cannot even be
made to be bounded from  below.)
   Thus the Belavin-Polyakov  instantons must
   appear as non-topological classical
solutions in the dual theory, where their stability
 is not automatically assured.
On the other hand, here we  show that
our  dual action agrees on shell
 with the action of the $O(3)$ nonlinear $\sigma-$model.  (In the
appendix we generalize this result to  $G/H$-models.)
It then follows that, on shell, the dual Euclidean action is bounded
from below, and that  classical solutions of the dual theory satisfy the Bogomol'nyi bound.
It also leads to a {\it dynamical} flux quantization condition.
For answering the question in the first paragraph,
a semiclassical path integral can be computed in the dual theory
 and compared with analogous calculations
 for the   $O(3)$ nonlinear $\sigma-$model.  (For example, one can
try to reproduce the results of Fateev,
Frolov and Schwarz\cite{ffs} in the dual theory.)
  We plan to report on these calculations in a forthcoming article.

In this letter,
 we give an explicit construction of the  instantons
of the dual theory.\footnote{For dual instantons in gravity see
 \cite{dig}.}     The construction involves
gluing Belavin-Polaykov  instantons   together with  corresponding
 anti-instantons   of opposite winding number.  The dual instantons
        are seen to have one  zero mode  which is
not present for Belavin-Polaykov instantons.

We first review the   $O(3)$ nonlinear $\sigma$-model,
 which  we shall refer to as the primary theory,
 and its dual formulation\cite{prwo}.
The target space for the    $O(3)$ nonlinear $\sigma$-model is  $S^2$,
which is span by  the  fields $\psi^i(x)$, $i=1,2,3$,
 $\psi^i(x)\psi^i(x) =1$.   We shall specialize to two-dimensional
  Euclidean space-time.     The standard  Lagrangian density $L$ is
\be L=\frac\kappa{2} \partial_\mu \psi^i
   \partial_\mu\psi^i \;,\label{orLa}
\ee
where  $\kappa$ is the coupling constant.
This system can also be re-expressed
in terms of $SU(2)$-valued fields $g(x)$.\cite{bgm},\cite{book}
  We take $g$ to be in
the defining representation and write
\be      \psi^i(x)\sigma^i= g(x)\sigma^3 g(x)^ {-1} \;,\label{psitg}\ee
     $\sigma_i$ being the Pauli matrices.
This introduces an additional $U(1)$ gauge degree of freedom, associated
with   $ g(x) \rightarrow g(x)\; \exp{-\frac i2\lambda(x)
 \; \sigma^3}\;.$  The corresponding    $U(1)$ connection is
 \be       A_\mu = i\;{\rm Tr} \sigma_3 (g^{-1}\partial_\mu g)\;.
\label{defA}      \ee
In addition, one can introduce a complex current
 \be \Pi_\mu   =  i\; \epsilon_{\mu\nu} {\rm Tr} \sigma^+
  ( g^{-1}\partial_\nu g) \;,\qquad\sigma^+ = \sigma^1 + i \sigma^2\;,
   \label{defpi} \ee  which
   rotates in the complex plane under a gauge transformation.
    The gauge invariant Lagrangian may be re-expressed in terms
     of these currents as
\be  L = \frac{\kappa} {2} |\Pi_\mu|^2    \;.
\label{stLagr}\ee
The equations of motion resulting from variations of $g$, $\delta g=
-\frac i2 g \sigma^i\epsilon^i$,    $\epsilon^i$ being infinitesimal,
state that the covariant curl of $\Pi$ is   zero:
\be \epsilon_{\mu\nu} D_\mu \Pi_\nu =0   \;.
\label{o3eom}\ee
where   the covariant derivative is defined by
$ D_\mu\Pi_\nu=\partial_\mu\Pi_\nu +i A_\mu \Pi_\nu\;.$
Along with the equations of motion (\ref{o3eom}),
 we have   three identities.  These are just the Maurer-Cartan equations,
which in terms of   $A_\mu$ and $ \Pi_\mu$, are
\beqa
  D_\mu\Pi_\mu    & =&0\;,                         \label{codiv} \\      F& =&
  -{i\over 2} \epsilon_{\mu\nu} \Pi_\mu\Pi_\nu^*\;,\label{do3eom} \eeqa
$F$ being the $U(1)$ curvature,
$ F=\epsilon_{\mu\nu} \partial_\mu A_\nu \;.  $
Finite action requires that we identify the points at infinity and
 compactify the Euclidean space-time to $S^2$.  The configuration space
is then a union of disconnected sectors associated with $\Pi_2(S^2)$.
  $F$ is  proportional to the winding number density
\be\rho=  \frac 1{8\pi}  \epsilon^{ijk}
 \epsilon_{\mu\nu} \psi^i\partial_\mu \psi^j\partial_\nu \psi^k
   \;,\label{instfld}\ee    and the  total flux is
  \be \int_{S^2} d^2x\; F =4\pi n \;,\label{qc}\ee
  $n$ being the winding number.

The dual Lagrangian $\tilde L$ in Minkowski space was specified
in \cite{prwo}.  It was given  in terms of a complex
  scalar field  $\chi$
   and the $U(1)$ connection $A_\mu$ (now regarded
   as  independent field variables).
   It was useful to also introduce an auxiliary scalar $\theta$.
One can perform a Wick rotation to obtain the corresponding Euclidean
action.  We specify the Wick rotation  later, and for now just
 assume the
Euclidean Lagrangian $\tilde L$  to have the general form
 $$ \tilde L =  \tilde L_0 + L_{BF} \;,$$
\be    \tilde L_0 = {\alpha\over 2} |D_\mu \chi|^2+ {{i\beta}\over 2}
\epsilon_{\mu\nu} (D_\mu \chi) (D_\nu \chi)^*
\;, \qquad    L_{BF} = \theta F \;.\label{totdL}\ee
 $\alpha\delta^{ab}$ and $\beta\epsilon^{ab}$ represent the dual
  metric and antisymmetric tensor, respectively.
The covariant derivative $D_\mu \chi $  is defined by
$    D_\mu \chi      = \partial_\mu \chi +i A_\mu\chi\;. $
  Under  gauge transformations,
 \be
 \chi\rightarrow e^{-i\lambda(x)}  \chi   \;,\qquad
     A_\mu\rightarrow A_\mu +
  \partial_\mu\lambda           \;,\label{dgt}\ee  while
$\theta$ is assumed to be gauge invariant, and thus so is $L_{BF}$.
Like in \cite{prwo}, we will assume
 that $\alpha$ and $\beta$ are independent of $\chi$ and $A$, and hence
    $\tilde L_0$ is gauge invariant.
 On the other hand, we  allow for  a nontrivial
 dependence on $\theta$.
    The expression for these functions is given below.
     For the dual theory to correspond
   to the primary theory, we should compactify the Euclidean space-time
 manifold to   $S^2$.  Then, in general, $A_\mu$ is not globally
defined, i.e. the  curvature two-form  is closed but not exact.

Following  \cite{prwo}, it is easy to show that we recover the
equations of the primary theory, i.e.,     (\ref{o3eom}),
      (\ref{codiv}) and (\ref{do3eom}),
 starting from the dual Lagrangian  (\ref{totdL}), for a certain $\alpha$
and $\beta$.  Furthermore,
although $\tilde L$ is not positive definite, we  show that
$\tilde S=\int_{S^2}d^2x\; \tilde L\ge 0 $  on shell, and
moreover that its numerical value is identical to that of  the primary
 Euclidean action  $ S=\int_{S^2} d^2x\;  L $.

             We first reproduce the equations  (\ref{o3eom}),
 (\ref{codiv}) and (\ref{do3eom}).
 For the equations of motion resulting  from
       variations of    $\chi$, we can ignore the $BF$-term.
From the assumption that $\alpha$ and $\beta$ are independent of $\chi$,
  we easily recover (\ref{codiv}), once we  define
   $\Pi_\mu$    according to
\be  \Pi_\mu   = -\alpha D_\mu \chi+i\beta \;
\epsilon_{\mu\nu}  D_\nu \chi \;. \label{ddefpi}
\ee
 This definition
leads to the identity \be {\rm Im} \; D_\mu \chi\; \Pi_\mu^* =0 \;,
\label{ident}\ee
and upon using the equations of motion  (\ref{codiv}), it follows that
$\epsilon_{\mu\nu}\; {\rm Im}\;   ( \chi\; \Pi_\nu^* )\; dx^\mu $
is a closed
one-form.   From   variations of  $A$ in $\tilde L$, it is also exact:
\be
  \epsilon_{\mu\nu}\partial_\nu\theta  =
 -{\rm Im}\;\chi\;\Pi_\mu^*\;. \label{defchi1}  \ee
Variations of $\theta$ in $\tilde L$  lead to
\be F\;= -{{\alpha'}\over 2} |D_\mu \chi|^2 -
 {{i\;\beta'}\over 2} \epsilon_{\mu\nu}
 (D_\mu \chi) (D_\nu \chi)^* \;,\label{fthord}\ee  the prime
indicating a derivative with respect to $\theta$.
This agrees with (\ref{do3eom}) provided that
\be \alpha'=2\alpha \beta\;
 \;,\qquad \beta'=(\beta^2+\alpha^2)\; \;.\ee
    These equations are solved by
\be \alpha=-\frac\kappa{\kappa^2-\theta^2}\;,\qquad
 \beta=\frac{\theta}{\kappa^2-\theta^2} \;,
\label{defab}\ee   up to a constant translation in $\theta$.
Eq. (\ref{fthord}) is then a fourth order equation for $\theta$ which, in
principle, can be used to eliminate the auxiliary scalar field.
$\kappa$  denotes the coupling constant of the dual theory.
From the Hamiltonian analysis of the Minkowski formulation of this system,
it is identical to the coupling constant $\kappa$ of the primary
theory.
   It remains to obtain  (\ref{o3eom}).
For this we need another identity, which is obtained by
   inverting (\ref{ddefpi}), using (\ref{defab}),
   to solve for $D_\mu \chi$:
\be  D_\mu \chi = \kappa\Pi_\mu -i\theta \epsilon_{\mu\nu}
\Pi_\nu  \label{dchi}                        \;.\ee
 Now   take the covariant curl to get
$ -i\kappa\;\epsilon_{\mu\nu} D_\mu \Pi_\nu =
F\; \chi -   \partial_\mu\theta\; \Pi_\mu -\theta\; D_\mu \Pi_\mu
  \;. $
The right hand side  vanishes upon imposing the equations of
motion  (\ref{codiv}),  (\ref{do3eom}) and  (\ref{defchi1}), and hence we
recover the equation of motion of the primary formulation (\ref{o3eom}).

 By comparing (\ref{totdL}) with the  Lagrangian in \cite{prwo}
[where we assume the metric tensor diag$(1,-1)$],
we see that the Wick rotation from Minkowski to Euclidean space-time
  affects scalar as well as vector fields:
$$ \partial_0 \rightarrow  i\partial_0   \;,\qquad
   A_0 \rightarrow  iA_0 \;,            $$
\be  \theta \rightarrow  i\theta   \;,\qquad
  \chi \rightarrow i\chi \;, \qquad      \chi^* \rightarrow i\chi^* \;.
 \ee
(We also added a total divergence to the Lagrangian in \cite{prwo}.)

The dual  Lagrangian (\ref{totdL}) can be re-expressed in several curious
ways.   One way is to substitute the definition of
$ \Pi_\mu $ in (\ref{ddefpi}) back into $\tilde L_0$ and integrate
by parts to get      \be \tilde L_0 =
\chi^*D_\mu \Pi_\mu  -\partial_\mu (\chi^* \Pi_\mu) \;. \label{form1}\ee
 Reality follows from (\ref{ident}).    The second term gives no
  contribution to the action for the domain $S^2$.  (For this note that
   $\chi^* \Pi_\mu$ is globally defined.)
   Moreover, the first  term, and hence  the action
   $\tilde S_0=\int_{S^2}d^2x\;\tilde L_0\;,$
vanishes when evaluated  on the space of classical solutions, which
we denote by $\tilde S_0|_{cl}=0$.       Alternatively,
we can write $\tilde L_0$ quadratically in terms of the currents
if we substitute  (\ref{dchi}) back into (\ref{totdL}),
\be \tilde L_0 = - \frac{\kappa} {2} |\Pi_\mu|^2   -
 {{i\theta}\over 2} \epsilon^{\mu\nu}  \Pi_\mu\Pi_\nu^*
\;.\label{Delta} \ee
  The first term is minus
 the primary Lagrangian (\ref{stLagr}) upon    identifying
the coupling constants of the theory.  Furthermore, the second term
is equivalent to the $BF$-term  after using (\ref{do3eom}).
Other possible forms for $\tilde L_0$ are obtained by taking
linear combinations of (\ref{form1}) and (\ref{Delta}).  Taking
 twice (\ref{form1}) minus (\ref{Delta}) gives
\be \tilde L =  \frac{\kappa} {2} |\Pi_\mu|^2   +
 \theta(F+{{i}\over 2} \epsilon^{\mu\nu}  \Pi_\mu\Pi_\nu^*)
+2\chi^*D_\mu \Pi_\mu  -2\partial_\mu (\chi^* \Pi_\mu) \;,\ee
where we added the $BF$-term.
 This implies that the primary and dual actions coincide on shell,
\be   \tilde S|_{cl}=\int_{S^2} d^2x\; \tilde L\;\bigg|_{cl} =
 \frac{\kappa} {2}  \int_{S^2} d^2x \; |\Pi_\mu|^2\;\bigg|_{cl} =
    S|_{cl}\;,\label{eqac}\ee       and thus
the dual action evaluated on the space of classical solutions is
positive definite (with the vacuum solution corresponding to vanishing
currents  $\Pi_\mu$.)             The result that
a dual action can be found which agrees on shell with the primary action
can be generalized to  $G/H$ models
for any Lie groups $G$ and $H\subset G$ (see Appendix).
 The dual action is (in Minkowski
space-time) is given in \cite{prwo}.

Although the space of field configurations in
 the dual version of the  $O(3)$ nonlinear $\sigma-$model
  is topologically
trivial, (\ref{eqac}) implies that the subspace of all
 classical solutions with
  finite Euclidean action is a union  disconnected regions.  The latter
are classified by the total flux, which we know from the primary theory
is quantized according to (\ref{qc}).  We can therefore say that the
 quantization  condition is  dynamically generated.
It  does not appear to result from any kinematic considerations of
  the  dual theory, as, classically, all values of the flux are
  allowed.\footnote{In this regard, note that  if the value of $\alpha$
 at spatial infinity
is restricted to being finite,  a bounded
 Euclidean action does not necessarily imply
 that $A$ must go to a pure gauge   at spatial infinity.}
    On the other hand, a
  semiclassical argument based on Wilson loops $W(C)=\exp{i\int_C A}$
 gives flux quantization, but it differs from (\ref{qc}).  Demanding
that the expectation value of $W(C)$ is independent of the coordinate
patch chosen on $S^2$ for any closed path $C$ gives
$\int_{S^2} d^2x\; F =2\pi\;\times$ integer.
 With this quantization condition, which is identical to the Dirac
quantization of magnetic charge,   we
can allow for, say, merons.\cite{Grs}
  However, such solutions are known to have  infinite  Euclidean action.

 The instantons and anti-instantons of Belavin and
Polyakov\cite{BP}     are  self-dual  and anti-self-dual   solutions,
respectively,  and they correspond to the
 minima of the Euclidean action of the primary theory
in every topological sector.  They are therefore
 `topologically' stable.
  For this one can  write  $L$ in (\ref{stLagr})  according to
\be L=\frac\kappa 4
 |\Pi_\mu \pm i\epsilon_{\mu\nu}\Pi_\nu |^2 \pm
 {{i\kappa}\over 2} \epsilon_{\mu\nu} \Pi_\mu\Pi_\nu^* \label{sdeq}\ee
 The Bogomol'nyi bound\cite{Bog} for the Euclidean action
 of the primary theory then follows from (\ref{do3eom})
 $$S =\int_{S^2} d^2 x\; L\ge 4\pi\kappa|n|\;,$$
 with the lower bound saturated by self-dual (instanton)
  configurations, i.e.
 $\Pi_\mu - i\epsilon_{\mu\nu}\Pi_\nu =0 $ when $n > 0 $, and
  anti-self-dual (anti-instanton) configurations, i.e.
 $\Pi_\mu + i\epsilon_{\mu\nu}\Pi_\nu =0 $ when $n < 0 $.
Of course, the  instantons (and anti-instantons) are
 also solutions of the dual
theory, and from (\ref{eqac}) they have the same value for the action
 as in  the primary theory, i.e. $\tilde S|_{cl}=4\pi\kappa |n|\;.$
However, $n$ cannot
 represent a topological index in the dual theory,
as the target space topology is trivial, and now
 stability  cannot be assured from topology.

Below we construct
the general form of the instanton solutions  in the dual theory.

We first review
 the construction of the most general instanton solutions
  in the primary theory.\cite{BP}  For this it was found convenient to
perform a stereographic projection, and write the scalar fields $\psi^i$
in terms of a complex function $W(x)$,
\be \psi^1+i\psi^2= \frac{2W} {1+|W|^2} \;,\qquad
   \psi^3  = \frac{|W|^2-1}{|W|^2+1}\;. \ee
In terms of this function the Lagrangian (\ref{orLa})
and instanton number density  (\ref{instfld})
 become
\be L=\frac{4\kappa}{(1+|W|^2)^2} (|\partial_zW|^2 + |\partial_zW^*|^2 )
\ee
\be \rho=\frac{1}{\pi(1+|W|^2)^2} (|\partial_zW|^2 - |\partial_zW^*|^2)
\ee
where we use the complex coordinate   $z= x_0+ix_1$.  From (\ref{sdeq})
instantons require that $L=4\pi\kappa\rho$.  This is only possible
for  $ \partial_{z^*}W =0$, and
therefore $W$ is  an analytic function of $z$.
Alternatively, anti-instantons require that $L=-4\pi\kappa\rho$,
 leading to
 $W$  being an analytic function of $z^*$.
For the choice of boundary conditions
$W\rightarrow 1$,    as $|x|\rightarrow \infty$,
the general instanton solution with winding number $n$
 has the  form\cite{BP}
\be W(z) =\frac{\prod_{i=1}^n (z-a_i)}{\prod_{j=1}^n (z-b_i)}
\;, \label{bp}\ee
where $a_i$ and $b_i$ are complex constants.

 To write down the currents $\Pi_\mu$ and connection one form $A$
  associated with the instanton,
 we must fix a gauge $g(W)$
for the $SU(2)$-valued field $g$.  In general, this can only be
done locally.
  A  gauge choice which is everywhere valid away from  the poles
 of $W(z)$ is  \be
g_S(W) =\frac1{\sqrt{1+|W|^2}}\pmatrix{W^* & -1\cr 1 & W}\;.
 \label{gsw} \ee
Alternatively, one that is every valid away from the zeros
 of $W(z)$ is  \be
g_N(W) =\frac{|W|}{\sqrt{1+|W|^2}}\pmatrix {1 & -W^{-1}\cr {W^*}^{-1} &
1}\label{gnw}\;. \ee    Since the general solution (\ref{bp}) contains
 poles as well as zeros,
 we will need to cover $S^2$ with at least two open regions,
  $R^2_S$ containing  the zeros and   $R^2_N$ containing  the poles.
We can then make the  gauge choice (\ref{gsw}) for $R^2_S$, and
(\ref{gnw}) for $R^2_N$.\footnote{On the other hand, a global gauge
exists for solutions containing only zeros, or only poles.  This will
require that $W$ have boundary value $\infty$ or $0$, respectively,
as $|x|\rightarrow \infty$.}
In   $R^2_S$,  we have the left invariant  one form
\be g_S^{-1} d g_S = \frac 1{1+|W|^2} \pmatrix{
\frac12  (WdW^* - W^* dW) & dW \cr - dW^* &
-\frac12  (WdW^* - W^* dW)   }\;,\label{mcfs} \ee
while in   $R^2_N$ the left invariant one form
$ g_N^{-1} d g_N$ is obtained by simply replacing $W$ by $-W^{-1}$
everywhere in (\ref{mcfs}).
The $z$ components of
the currents and $U(1)$ connection   are then
\beqa    \Pi_z^{(S)} &=&  \frac 12 (\Pi_0-i\Pi_1 )   =0 \cr
       \tilde  \Pi_z^{(S)}& =& \frac 12 (\Pi_0^*-i\Pi_1^* ) =
 \frac{-2\;\partial_zW}        {1+|W|^2}     \cr
  A_z^{(S)} & = &\frac 12 (A_0 - i A_1)= - i\; \partial_z \;
   {\rm ln}\; {(1+|W|^2)}   \;,
 \label{tppa}                \eeqa
 in   $R^2_S$, and
\beqa    \Pi_z^{(N)} &=&  \frac 12 (\Pi_0-i\Pi_1 )   =0 \cr
       \tilde  \Pi_z^{(N)}& =& \frac 12 (\Pi_0^*-i\Pi_1^* ) =
 \frac{2\;\partial_zW^{-1}}        {1+|W|^{-2}}     \cr
  A_z^{(N)} & = &\frac 12 (A_0 - i A_1)=
 - i\; \partial_z \;
      {\rm ln}\; {(1+|W|^{-2})}
 \label{tppab}                \eeqa               in   $R^2_N$.
 In both (\ref{tppa}) and (\ref{tppab})  we used  $ \partial_{z^*}W =0$.
As stated earlier this is consistent with the condition of
 self-duality, i.e., $\Pi_z=0$.
It is  easy to write down the transition function $\lambda^{(NS)}$
 between
the two gauges in the overlapping region $R^2_S \cap R^2_N$:
 \be  \lambda^{(NS)}= i\;{\rm ln}\; \frac {W(z)}{W(z)^*}  \label{trfn}
   \ee
which transforms $A_z$ and $\tilde\Pi_z$ according to
$ A_z^{(N)}  =     A_z^{(S)} +  \partial_z\;\lambda^{(NS)} $ and
$    \tilde\Pi_z^{(N)}  = e^{i\lambda^{(NS)}} \; \tilde \Pi_z^{(S)}  $
(The above analysis can easily be
repeated for anti-instantons, corresponding to   $ \tilde  \Pi_z=0$.)

Before writing  down the instanton solution in the dual theory, we first
 look at the implications of self-duality and anti-self-duality.
Eqs. (\ref{defchi1}) and (\ref{dchi})  for the scalar fields,
  can be expressed as
\beqa
 \partial_z\theta & =& \frac 12 (\chi \tilde \Pi_z - \chi^* \Pi_z )\cr
D_z \chi & =&\partial_z\chi + iA_z \chi= (\kappa+\theta)\Pi_z \cr
D_z \chi^*&=&\partial_z\chi^* - iA_z \chi^*=
 (\kappa-\theta)\tilde\Pi_z
\label{deqct}           \eeqa
  Instantons, i.e. $\Pi_z=0$,  imply $D_z \chi=0$, while
anti-instantons, i.e. $\tilde \Pi_z=0$,  imply $D_z \chi^*=0$.
In either case, we can then write the connection in terms of
scalar fields.   Furthermore,      from (\ref{deqct}), it follows that
  $(\theta - \kappa)^2 +|\chi|^2 $ is a constant for $\Pi_z=0$,
  while
  $(\theta + \kappa)^2 +|\chi|^2 $ is a constant for
   $\tilde\Pi_z=0$.
Therefore, when the currents are restricted to being
 self-dual or anti-self-dual, the
 scalar fields of the dual theory define a two-sphere, in analogy
with the scalar fields of the primary theory.   [One major
difference with the
primary theory, though, is that while $\psi^i$ are gauge invariant,
$\chi$ is not.
 $\chi$ contains only one gauge invariant degree of freedom, and hence
the gauge invariant degrees of freedom in the
 self-dual or anti-self-dual fields  of the dual theory, in fact,
 span  $S^1$.   As $\Pi_2(S^1)=0$, the topology of this space
 is  trivial.]   We can  parametrize the scalar fields
in terms of a complex function, say $V(x)$, via a stereographic projection,
 as was done for the primary theory.          In the case of instantons,
i.e., $\Pi_z=0$, we write
\beqa \chi& =&\frac{2RV^*}  {1+|V|^2} \;,\cr
\theta & =&R \; \frac{1-|V|^2}{1+|V|^2}\;     +\kappa  \;,
\label{sfct} \eeqa
where $R$ is the radius of the two-sphere.   This expression is valid
in any open subset of $S^2$.
   By comparing $D_z \chi=0$ with the equations of motion
$D_z (\tilde\Pi_z)^*=0$ , we get that $(\tilde\Pi_z)^*=G(z)^*\;\chi$.
$G$ is an  analytic function of $z$, and from the last equation in
(\ref{deqct}), it is equal to $-\frac 1R\partial_z \ln |V|^2 $.
Then  up to a phase (which
can be gauged away) $V$ is either analytic or anti-analytic in $z$.
In general, both cases are  needed for the global description
of solutions with non-zero total flux.   The global description
is obtained by matching solutions in the overlapping regions of
 different open subsets of $S^2$.
An easy way to proceed is to use our previous result that
 finite action solutions of the dual theory correspond to
 finite action solutions of the primary theory.
By identifying  the currents and connections of the primary theory
(\ref{tppa}) and (\ref{tppab}) with those derived from (\ref{sfct}),
we get the gauge choice
$V(z)=W(z)$   in $R^2_S$, and $V(z^*)=-1/W(z)^*$ in $R^2_N$.
  The transition function is once again
given by  (\ref{trfn}).
The integration constant $R$ drops out of the expression for the currents
and connections, and hence represents a degeneracy in the space
 of solutions in the dual theory.   This implies that the
dual instantons have  a zero mode which is not present for
the instantons in the primary theory.       In matching the solutions
for $\chi$ and $\theta$ in $R^2_S$ and  $R^2_N$,
we note that $\theta$ is gauge invariant.  Then if we set
$V(z)=W(z)$   in $R^2_S$, and $V(z^*)=-1/W(z)^*$ in $R^2_N$
in  (\ref{sfct}) we must switch the sign of $R$ in the two regions.
The dual instanton solution is thus
\beqa  \chi(z) &= &
-\frac{2RW(z)^*}  {1+|W(z)|^2} \quad {\rm in}\;R^2_S\;  ,  \cr
&=&-\frac{2RW(z)}  {1+|W(z)|^2} \quad {\rm in}\;R^2_N \;, \eeqa
\be\theta (z)  =R \; \frac{|W(z)|^2-1}{|W(z)|^2+1}\;     +\kappa  \;.
 \ee         If
 $(\psi^1_{(n)},\psi^2_{(n)},\psi^3_{(n)})$ corresponds to the
  $n-$instanton solution of the primary theory
 expressed in terms of the  fields $\psi^i$,
  then the dual   $n-$instanton solution
can be written
 \be ( \frac{\chi^1}R,  \frac{\chi^2}R, \frac{\theta -\kappa}R )
 =\left\{\matrix{(-\psi^1_{(n)},\psi^2_{(n)},\psi^3_{(n)})
   & {\rm in}\;R^2_S \cr
                (-\psi^1_{(n)},-\psi^2_{(n)},\psi^3_{(n)})
                  & {\rm in}\;R^2_N }\right.
                \;,\label{gia}\ee      where $\chi=\chi^1+i\chi^2$.
  Thus  instantons in the dual theory are obtained by gluing
 instantons  of the  primary theory  together with  anti-instantons
of opposite winding number, the latter being
 obtained by switching the
  orientation of one of the components.

An analogous result can be found for the anti-instantons of
 the dual theory.  In that case, the right hand side of (\ref{gia})
gets replaced by
 $( \frac{\chi^1}R,  \frac{\chi^2}R, \frac{\theta +\kappa}R )$.

\bigskip
The authors are grateful to  D. O'Connor for useful discussions.

\bigskip

{\bf Appendix}

Here we show that for any nonlinear  $G/H$ model,
  the dual action \cite{prwo}  agrees on shell with the
primary action, up to boundary terms.

 Say $G$ and $H\subset G$ are $N$ and $N-M$ dimensional groups,
respectively, with
the former generated by $e_i, \;i=1,2,..N$, and
 having commutation relations:
$  [e_i,e_j]=c_{ij}^k e_k        $
We can split the generators into  $e_a,$ $a=1,2,...M$ and
  $\hat{e}_\alpha=e_{M+\alpha},$ $\alpha=1,2,...N-M$,
   the latter generating $H$,
$    [\hat{e}_\alpha,\hat{e}_\beta]=\hat{c}_{\alpha \beta}^\gamma
\hat{e}_\gamma          \;  ,$
$  \hat{c}_{\alpha\beta}^\gamma =c_{M+\alpha
  \;\;M+\beta}^{M+\gamma  }    \;.$
 We will  assume
that the metric ${\tt g}_{ij}$ on $G$ is nondegenerate and block
diagonal, i.e.
$     {\tt g}_{a\; M+\alpha}  =0 \;.\label{metr}$
The structure constants satisfy
$ c_{M +\alpha\;M +\beta}^c =0\;,$       $
c_{M +\alpha\;b}^{M +\gamma}   =0 \;.$   The
second relation follows
 from the first, using  the invariance property
$c^i_{jk}{\tt g}_{i\ell} ={\tt g}_{ji}c^i_{k\ell}$.

In the primary theory,  the fundamental fields   $g(x)$    have values
 in $G$.
Utilizing the group metric  ${\tt g}_{ij}$ projected onto
 $G/H$, the primary Lagrangian can be expressed as
\be L=        -
\frac\kappa {2} {\tt g}_{ab} ( g^{-1}\partial_\mu g)^a
 (g^{-1}\partial^\mu g)^b
 \;,  \label{ngstLagr}\ee
 where $a,b=1,2,...M$ and    $\kappa$ is the coupling constant.
$( g^{-1}d g)^a  $  denotes the $e_a$ component of the one-form $
 g^{-1}d g $.     $L$  is  gauge invariant under
 $  g(x)\rightarrow g(x) h(x)\;, \quad h(x)\in H\;,$
and consequently defines a theory on $G/H$.  $L$ is also
 invariant under global transformations
 $ g\rightarrow g_0g       \;,\quad g_0 \in G  \;.$
  There are now
$M$ equations of motion resulting from variations of $g$, and they
    can  be written as
\be \epsilon^{\mu\nu}(D_\mu \pi_\nu)^a =0   \;,  \label{nao3eom}\ee
  where
  \be \pi_\mu ^ a  =\epsilon_{\mu\nu}
 ( g^{-1}\partial^\nu g)^a \;,\qquad
  A^\alpha_\mu = (g^{-1}\partial_\mu g)^{M+\alpha}\;.\label{gdefA}
    \ee    and
the covariant derivative is now defined by
$ (D_\mu\pi_\nu)^a=\partial_\mu\pi^a_{\nu} + c^a _{M+\beta\;\; c}
 A_\mu^\beta  \pi_{\nu}^c\;. $
 $A_\mu^\alpha$ transforms as components of an $H$ connection one-form.
In addition to the equations of motion (\ref{nao3eom}),
 we have  $N$ Maurer-Cartan equations:
\beqa
 (D^\mu\pi_\mu)^a & =& \frac12 c^a_{bc}  \epsilon^{\mu\nu}\pi_\mu^b
 \pi_\nu^c\;,\label{dngcodiv} \\
F^\alpha & =&{1\over 2}c_{bc}^{M+\alpha}  \epsilon^{\mu\nu}
\pi_\mu^b\pi_\nu^c\;.\label{gndo3eom} \eeqa  Now, in general,
  the covariant
divergence of $\pi_\mu^a$ need not vanish.
$F^\alpha$ is the $H$ curvature,
$ F^\alpha=\epsilon^{\mu\nu}( \partial_\mu A_\nu^\alpha
  +\frac12 \hat{c}
^\alpha_{\beta\gamma}A^\beta_\mu A^\gamma_\nu)\;      \;.$
In terms of the currents  $\pi_\mu ^ a $, $L$ can be written
\be L=
\frac\kappa {2} {\tt g}_{ab} \;\pi_\mu ^ a      \pi^{\mu  a}
 \;.  \label{npstLagr}\ee

The dual action  (in Minkowski
space-time) is given in \cite{prwo}.
It is expressed in terms of $N$ scalar fields,
$\chi_a$ and
$\theta_\alpha$, along with the Yang-Mills connection one form
 $A^\alpha$,
which undergo gauge transformations   \beqa
 \delta\chi_a& =&c^b_{M+\alpha\;a}\lambda^\alpha  \chi_b \label{gtochi} \\
\delta \theta_\alpha & =&\hat{c}^\gamma_{\beta \alpha}\lambda^\beta
  \theta_\gamma                                      \label{gtpthe} \\
\delta A^\alpha & =&d\lambda^\alpha+ \hat{c}^\alpha_{\beta \gamma}
A^\beta \lambda^\gamma  \;,\label{gtovp}   \eeqa
    $\lambda =\lambda^\alpha
\hat{e}_\alpha$ being
an infinitesimal element of the Lie algebra ${\cal H}$
 of $H$.
The   Lagrangian density is
\be \tilde L =   -{1\over 2}{\alpha}^{ab} (D_\mu \chi)_a
 (D^\mu \chi)_b-
 {1\over 2} \epsilon^{\mu\nu}\beta^{ab} (D_\mu \chi)_a (D_\nu \chi)_b
 -\theta_\alpha F^\alpha   \;,        \label{gendL} \ee
where  the covariant derivative of $\chi_a$ is defined according to
$ (D\chi)_a =d\chi_a +c^b_{a\;M+\alpha}A^\alpha \chi_b \;,$
and  the dual metric ${\alpha}^{ab}$   and
the antisymmetric tensor $\beta^{ab}$  are given by
\be \alpha=\biggl(\kappa{\tt g}-
\frac 1\kappa{\tilde f}  {\tt g}^{-1}
{\tilde f}\biggr)^{-1} \;,   \qquad
  \beta =-\frac 1\kappa {\tt g}^{-1}{\tilde f}
    \alpha \;,\label{tgss}\ee        where
\be   {\tilde f}_{ab}=
c^c_{ab}\chi_c +c^{M+\alpha}_{ab}\theta_\alpha
 \;,\label{tfsac}\ee  and ${\tt g}$ is the group metric
 projected onto $G/H$.     $\alpha$ in (\ref{tgss}) is
 symmetric by inspection, while
antisymmetry for $\beta$ follows after using the identity
$ {\tilde f} \alpha  {\tt g}=  {\tt g}\alpha {\tilde f}\;.$
Upon varying  $\chi_a$, $\theta_\alpha$ and $A^\alpha_\mu$ and
applying identities, one recovers  the
equations (\ref{nao3eom}), (\ref{dngcodiv}) and (\ref{gndo3eom}) of
the primary system \cite{prwo}.    For this
$\pi_\mu^a$ is now defined  by  \be  \pi_\mu^a   =
-\alpha^{ab}(D_\mu \chi)_b-\beta^{ab}
\epsilon_{\mu\nu}  (D^\nu \chi)_b \;. \label{gdefpi}
\ee   Substituting this expression back into $\tilde L$ and integrating
by parts gives
\be \tilde L =   -{1\over 2} \chi_a \; (D_\mu \pi^\mu)^a
 + {1\over 2} \partial_\mu ( \chi_a \pi^{\mu a})
 -\theta_\alpha F^\alpha   \;.        \label{gendL1} \ee
One can also write
\be \tilde L =   -{\kappa\over 2} {\tt g}_{ab} \; \pi^{\mu a}\pi^b_\mu
 - {1\over 2} \tilde f_{ab}\;\epsilon^{\mu\nu}  \pi^{a}_\mu\pi^b_\nu
 -\theta_\alpha F^\alpha   \;.        \label{gendL2} \ee
Finally, twice (\ref{gendL1})  minus (\ref{gendL2}) gives
\beqa \tilde L &=&
   {\kappa\over 2} {\tt g}_{ab} \; \pi^{\mu a}\pi^b_\mu
    +\partial_\mu ( \chi_a \pi^{\mu a})      \cr
& & +\;\chi_c\; \biggl({1\over 2}
 c^c_{ab}\;\epsilon^{\mu\nu}  \pi^{a}_\mu\pi^b_\nu
 - (D_\mu \pi^\mu)^c \biggr )     \;
+\;\theta_\alpha\; \biggl({1\over 2}
c^{M+\alpha}_{ab}\;\epsilon^{\mu\nu}  \pi^{a}_\mu\pi^b_\nu
-  F^\alpha   \biggr)   \;.
      \label{gendL3} \eeqa
The second line vanishes after using the equations of motion
[which were the Maurer-Cartan equations
(\ref{dngcodiv}) and (\ref{gndo3eom})   in the primary theory].  Hence,
on shell, the dual action agrees with the primary action up to a
boundary term.

 \end{document}